\documentclass[journal]{IEEEtran}
\usepackage{amsfonts}
\usepackage{amssymb}
\usepackage{graphicx}
\usepackage{amsmath}
\usepackage{amsfonts,epsfig, multirow, floatflt,array}
\usepackage{epsfig,graphics,graphicx,latexsym,verbatim,slashbox}
\usepackage{subfigure}
\usepackage{cite}
\usepackage{color}
\usepackage{array}
\usepackage{booktabs}
\usepackage{multicol}
\usepackage{amsthm}
\usepackage{algorithm}
\usepackage{algpseudocode}
\usepackage{stfloats}
\usepackage{flushend}
\usepackage{float}
\usepackage{epstopdf}

\begin{document}
%
% paper title
% Titles are generally capitalized except for words such as a, an, and, as,
% at, but, by, for, in, nor, of, on, or, the, to and up, which are usually
% not capitalized unless they are the first or last word of the title.
% Linebreaks \\ can be used within to get better formatting as desired.
% Do not put math or special symbols in the title.
\title{Analysis of Proportional Fair Scheduling\\ Under Bursty On-Off Traffic}
%
%
% author names and IEEE memberships
% note positions of commas and nonbreaking spaces ( ~ ) LaTeX will not break
% a structure at a ~ so this keeps an author's name from being broken across
% two lines.
% use \thanks{} to gain access to the first footnote area
% a separate \thanks must be used for each paragraph as LaTeX2e's \thanks
% was not built to handle multiple paragraphs
%

\author{Fei Liu, Janne Riihij\"{a}rvi, and Marina Petrova
        %John~Doe,~\IEEEmembership{Fellow,~OSA,}
        %and~Jane~Doe,~\IEEEmembership{Life~Fellow,~IEEE}% <-this % stops a space
\thanks{Fei Liu and Janne Riihij\"{a}rvi are with the Institute for Networked Systems, RWTH Aachen University, 52072 Aachen, Germany (email: fei@inets.rwth-aachen.de; jar@inets.rwth-aachen.de).}% <-this % stops a space
\thanks{Marina Petrova is with KTH Royal Institute of Technology and the Institute for Networked Systems at RWTH Aachen University (email: mpe@inets.rwth-aachen.de).}
}

% note the % following the last \IEEEmembership and also \thanks -
% these prevent an unwanted space from occurring between the last author name
% and the end of the author line. i.e., if you had this:
%
% \author{....lastname \thanks{...} \thanks{...} }
%                     ^------------^------------^----Do not want these spaces!
%
% a space would be appended to the last name and could cause every name on that
% line to be shifted left slightly. This is one of those "LaTeX things". For
% instance, "\textbf{A} \textbf{B}" will typeset as "A B" not "AB". To get
% "AB" then you have to do: "\textbf{A}\textbf{B}"
% \thanks is no different in this regard, so shield the last } of each \thanks
% that ends a line with a % and do not let a space in before the next \thanks.
% Spaces after \IEEEmembership other than the last one are OK (and needed) as
% you are supposed to have spaces between the names. For what it is worth,
% this is a minor point as most people would not even notice if the said evil
% space somehow managed to creep in.

% The paper headers
\markboth{This is the author's version of an article that has been accepted for publication in IEEE Communications Letters.}%
{Shell \MakeLowercase{\textit{et al.}}: Analysis of Proportional Fair Scheduling Under Bursty On-Off Traffic}
% The only time the second header will appear is for the odd numbered pages
% after the title page when using the twoside option.
%
% *** Note that you probably will NOT want to include the author's ***
% *** name in the headers of peer review papers.                   ***
% You can use \ifCLASSOPTIONpeerreview for conditional compilation here if
% you desire.

% If you want to put a publisher's ID mark on the page you can do it like
% this:
%\IEEEpubid{0000--0000/00\$00.00~\copyright~2015 IEEE}
% Remember, if you use this you must call \IEEEpubidadjcol in the second
% column for its text to clear the IEEEpubid mark.

% use for special paper notices
%\IEEEspecialpapernotice{(Invited Paper)}

% make the title area
\maketitle

% As a general rule, do not put math, special symbols or citations
% in the abstract or keywords.
\begin{abstract}
Proportional fair scheduling (PFS) has been adopted as a standard solution for fair resource allocation in modern wireless cellular networks. With the emergence of heterogeneous networks with widely varying user loads, it is of great importance to characterize the performance of PFS under bursty traffic, which is the case in most wireless streaming and data transfer services. In this letter, we provide the first analytical solution to the performance of PFS under bursty on-off traffic load. We use the Gaussian approximation model to derive a closed-form expression of the achievable user data rates. In order to further improve the accuracy of our baseline analytical solution for multi-cell networks, we design a hybrid approximation by employing multi-interference analysis. The simulation results verify that our model guarantees extremely low data rate estimation error, which is further insensitive to changes in session duration, traffic load and user density.
\end{abstract}

% Note that keywords are not normally used for peerreview papers.
\section{Introduction}\label{SecIntro}
Opportunistic scheduling provides an effective mechanism to improve transmission performance by exploiting channel fluctuations in multiuser wireless communication networks~\cite{Lei2008}. Among various scheduling schemes, proportional fair scheduling (PFS) has been widely adopted since it provides an excellent balance between high throughput and user fairness~\cite{WCNC2008}. Thus performance analysis of PFS is important to provide guidelines for its optimization and application. In particular, analytical results can be used for admission control, radio resource management, network planning, and so on~\cite{ICC2014D}.
\par
There have been several related works in the literature which focus on the performance analysis of PFS under saturated traffic, where the active user set is static because users always have data to transmit. We can broadly classify the existing approaches into two major groups according to their analytical models of user data rates. The first group is designed for the single-cell network based on either symmetric relative fluctuation of user channels~\cite{CL2012} or Gaussian approximation of instantaneous data rates~\cite{WCNC2008}. The second group considers stochastic signal-to-interference-plus-noise ratio (SINR) distribution of each user in multi-cell networks and computes throughput with the conditional probability distribution of the scheduled SINR under PFS. The stochastic SINR model in \cite{SECON2015} is carefully designed using multi-interference analysis (MIA) and achieves more accurate estimation of user throughput in multi-cell networks compared to the single-cell analysis.
\par
Compared to the saturated traffic model, the bursty traffic model is more realistic especially for the analysis of dynamic service quality. It is important for modeling the increasing web usage which is becoming dominant in mobile networks nowadays. The packet-level performance has been investigated in \cite{ACMTran2005} by using processor sharing methodology. However, their analytical results are limited to the scenarios where the relative fluctuations of user channels are symmetric. On the other hand, high speed streaming services, such as video on demand (VOD) and video conferencing, have become more popular recently. The wireless transmission of these service types can be modeled as bursty on-off traffic at session level, which has not yet been considered for the performance analysis of PFS. Therefore, in this letter, we present the first analytical solution for the performance of PFS under bursty on-off traffic. We derive a closed-form expression of user data rates based on GA. In order to improve the accuracy of analytical results in multi-cell networks, we then design a hybrid approximation model by carefully combining GA and MIA approaches. We compare the analytical performance of PFS with the results obtained from simulations to verify the accuracy of our models.
\par
\section{System Model}
We consider a downlink network containing multiple base stations (BSs). We denote the BS in the considered cell as $b$ and the set of user terminals associated to it as
%We denote the set of the BS indices by
%\begin{equation}
%\label{EqBSSet}
%{\bf{B}} = \left\{ {b\left| {b = 1, \ldots ,B} \right.} \right\}.
%\end{equation}
%\par
%The index set of the user terminals is denoted by
\begin{equation}
\label{EqUserSet}
{\bf{U}}_b = \left\{ {u\left| {u = {1},2, \ldots ,{\left|{\bf{U}}_b\right|}} \right.} \right\}.
\end{equation}
\par
%\subsection{Channel Measurement}
In each transmitted frame, the BS distributes resource blocks (RBs) to the associated users with PFS~\cite{3GPP36211}. The PFS considered in this paper uses the data rate-based scheduling metric, i.e., the ratio between the instantaneous and long-term averaged user data rate. We assume flat fading channels such that the RBs within the considered bandwidth undergo identical Rayleigh fading. Without loss of generality, we focus on the performance analysis of PFS with one certain RB. Thus, the instantaneous received power of the reference signal (RS) at user $u$ from BS $b$ is modeled as
\begin{equation}
\label{EqRecPow}
{P_{u,{b}}} =  {p_b} {L_{u,{b}}}{\left\| {{h_{u,{b}}}} \right\|^2},
\end{equation}
\noindent where $p_b$ is the RS transmit power of BS $b$, ${L_{u,{b}}}$ is the channel gain of path loss and shadow fading, and $h_{u,b}$ is the normalized Rayleigh fading gain of user $u$ from BS $b$ which is modeled as a circularly symmetric complex Gaussian random variable with mean value 0 and covariance 1.
%The power gain of ${\left\| {h_{u,b}} \right\|^2}$ is exponentially distributed with a unit mean value.
Thus, $P_{u,b}$ is a random variable with the exponential distribution and its mean value is given as
\begin{equation}
\label{EqDefLambdaBS}
p_{u,b} = \mathbb{E}\left[P_{u,b}\right] =  {p_b}{L_{u,{b}}}.
\end{equation}
\noindent This can be estimated by the detection of RS-received-power (RSRP) which is the average power of the symbols that carry cell-specific RSs~\cite{3GPP36211}. A user reports RSRPs to its serving BS for channel quality detection. Thus, the BSs can make system decisions, such as for resource allocation and inter-cell handover, according to the reported information.
\par
The total instantaneous received RS power of user $u$ is
\begin{equation}
{P_u}{\rm{ = }}{P_{u,b}}{\rm{ + }}\sum\limits_{i \in {{\bf{I}}_u}} {{P_{u,i}}} {\rm{ + }}{\sigma _N},
\end{equation}
\noindent where ${\bf{I}}_u$ is the interfering BS set of user $u$, including $ \left| {{{\bf{I}}}_u} \right|$ independent inter-cell interferers, and ${\sigma _N}$ is the noise power. The mean value of the total received power is denoted as $p_{u}=\mathbb{E}\left[P_{u}\right]$, which can be estimated according to the received signal strength indicator (RSSI). The feedback of RSRPs and RSSIs from user terminals can be used to estimate the probability distributions of user SINRs~\cite{SECON2015}.
The instantaneous SINR is expressed as
\begin{equation}
\label{EqSIRBS}
{\Phi _u}{\rm{ = }}{{{P_{u,b}}} \over {\sum\limits_{i \in {{\bf{I}}_u}} {{P_{u,i}}} {\rm{ + }}{\sigma _N}}}.
\end{equation}
\par
%\subsection{Bursty On-off Traffic Model}
The data traffic is modeled as a semi-Markov on-off process that is assumed to be the independent and identically distributed (i.i.d.) among users. We assume that users can fully utilize the link capacity during session periods, and hence they always have data to transmit when they are active under \emph{on} states. We use Pareto distribution for modeling the duration of the \emph{on} state and exponential distribution for the duration of the \emph{off} state, which are denoted as follows~\cite{WWW2008}:
\begin{equation}\label{ondistr}
{F_{\rm{on}}}\left( d \right) = 1 - {\left( {\frac{{{\beta}}}{d}} \right)^\alpha },\quad d\ge\beta,
\end{equation}
\begin{equation}\label{offdistr}
{F_{\rm{off}}}\left( d \right) = 1 - \exp \left( {\lambda d} \right),\quad d > 0.
\end{equation}
\noindent Here parameters $\alpha$, $\beta$ and $\lambda$ decide the characteristics of the on-off duration distributions. Accordingly, the mean durations of the two states are given as
\begin{align}\label{onoffmean}
{D_{\rm{on}}}& = \frac{{\alpha \beta }}{{\alpha  - 1}}, \quad {\rm{and}} \\
{D_{\rm{off}}}& = {\lambda ^{ - 1}}.
\end{align}
\par
The traffic load is defined as the duty cycle of the on-off process, which can be calculated as
\begin{equation}\label{trafficload}
\rho  = \frac{{{D_{\rm{on}}}}}{{{D_{\rm{on}}} + {D_{\rm{off}}}}} = {\left[ {1 + \frac{{\alpha  - 1}}{{\alpha \beta \lambda }}} \right]^{ - 1}}.
\end{equation}
\noindent Specifically, when $\rho = 1$, the traffic in the system is saturated.
\par
\section{Analysis of PFS Using GA}
 Under bursty on-off traffic, the active user set is dynamic due to the constant changes of user states. However, it keeps steady within the session duration which is still relatively much larger than the RB scheduling period. Thus, we can calculate the achievable date rate of a user under every possible combinations of the active users and their corresponding probabilities. The average throughput of the user is the weighted sum of these data rates in terms of the probabilities, which is calculated as
%we derive the closed-form results of the estimated user data rate as
\begin{equation}\label{Rburst}
{{R}_u}\left( {{\bf{U}}_b,\rho } \right) = {r_u}\rho \sum\limits_{u \in \left({\bf{V}} \subseteq {\bf{U}}_b \right)} {\left[ {\frac{{{{G}_u}\left( {\bf{V}} \right)}}{{\left| {\bf{V}} \right|}}{\rho ^{\left| {\bf{V}} \right| - 1}}{{\left( {1 - \rho } \right)}^{\left| {\bf{U}}_b \right| - \left| {\bf{V}} \right|}}} \right]} ,
\end{equation}
\noindent where ${{r}_u}$ is the average data rate of user $u$ while it is scheduled alone, ${{ G_u}\left( {\bf{V}} \right)}$ is the PFS performance gain over the round-robin (RR) scheduling under saturated traffic when user $u$ is active along with other users in $\bf{V}$. When the total number of users is high, the computational complexity of \eqref{Rburst} is large since ${{ G_u}\left( {\bf{V}} \right)}$ is different for each subset $\bf{V}$ that satisfies $u \in\left( {\bf{V}} \subseteq {\bf{U}}_b\right)$. Therefore, it needs to be calculated independently for each of the $2 ^{ \left( {\left| {\bf{U}}_b \right| - 1} \right)}$ possible cases.
\par
%\subsection{Data Rate Estimation with Gaussian Approximation}
%\subsection{Closed-form Solution for PFS Performance}
With the aim of tractable analysis, ${{ G_u}\left( {\bf{V}} \right)}$ can be approximated so that the effect of $\bf{V}$ on ${{ G_u}\left( {\bf{V}} \right)}$ is only through the number of users in it. To this end, the instantaneous user data rates are modeled with the Gaussian distribution in~\cite{WCNC2008}. An alternative approach is assuming that the normalized rates are i.i.d and linear in SINR~\cite{ACMTran2005}. The former one is referred to as GA and is adopted in this paper due to its higher estimation accuracy for multi-cell networks, which is compared with the latter one below.
\par
The performance gain of PFS over RR under saturated traffic is estimated with GA as
\par
\begin{equation}\label{eqGAfullgain}
\tilde G_u \left( {\bf{U}}_b \right) %=& 1 + \frac{{{\sigma _{{u}}}}}{{{{ r}_u}}}\left\{ {\left| {\bf{U}} \right|\int_{ - \infty }^\infty  {z\frac{{{e^{ - {z^2}/2}}}}{{\sqrt {2\pi } }}} {{\left[ {{F_{\left( {0,1} \right)}}\left( z \right)} \right]}^{\left| {\bf{U}} \right| - 1}}dz} \right\} \notag\\
 = 1 + \frac{{{\sigma _{{u}}}}}{{{{  r}_u}}} { \int_0^1 z d{\left[ {{F_{\left( {0,1} \right)}}\left( z \right)} \right]^{\left| {\bf{U}}_b \right|}}},
\end{equation}
\noindent where ${{F_{\left( {0,1} \right)}}\left( z \right)}$ is the cumulative distribution function (CDF) of the standard normal distribution, $\sigma _{u}$ is the standard deviation of the user data rate. $r_u$ and $\sigma_u$ are calculated according to (3) and (4) in \cite{WCNC2008}.
\par
%%%%%%%%%%%%%%%%%%%%%%%%%%%%%%%%%%%%%%%%%%%%%%%%%%%%%%%%%%%%%%%%%%%%%%%%%%%%%%%%%%%%%%%%%%%%%
\begin{figure*}[t]
\centering
\begin{minipage}[t]{0.32\linewidth}
{\includegraphics[width=1.9in]{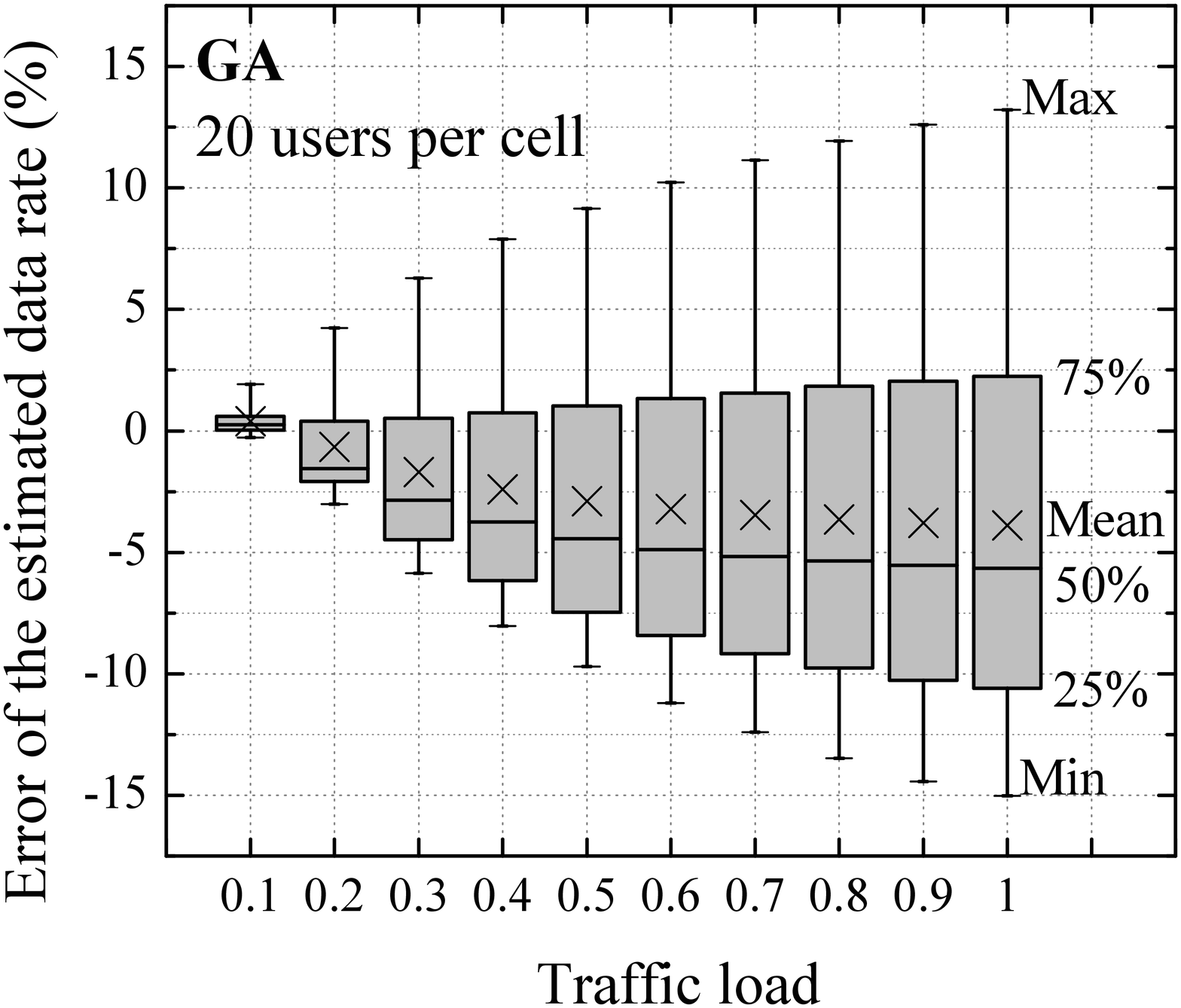}
 \caption{Error of the GA-based data rate \protect\\ estimation under bursty on-off traffic.}
\label{GA20UserBurst}}
\end{minipage}
\begin{minipage}[t]{0.32\linewidth}
{\includegraphics[width=1.9in]{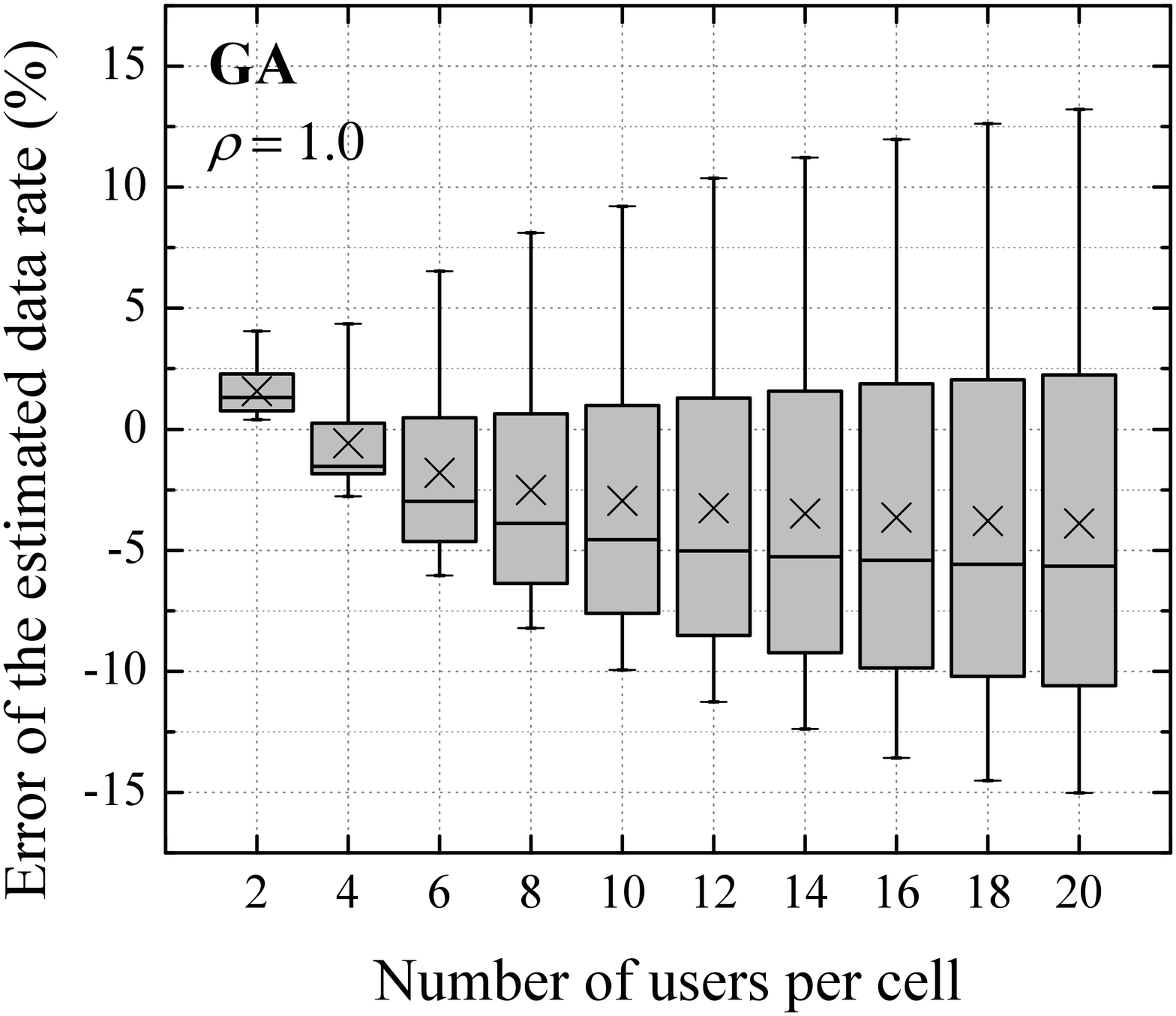}
 \caption{Error of the GA-based data rate \protect\\ estimation under saturated traffic.}
\label{GANUserFull}}
\end{minipage}
\begin{minipage}[t]{0.32\linewidth}
{\includegraphics[width=1.9in]{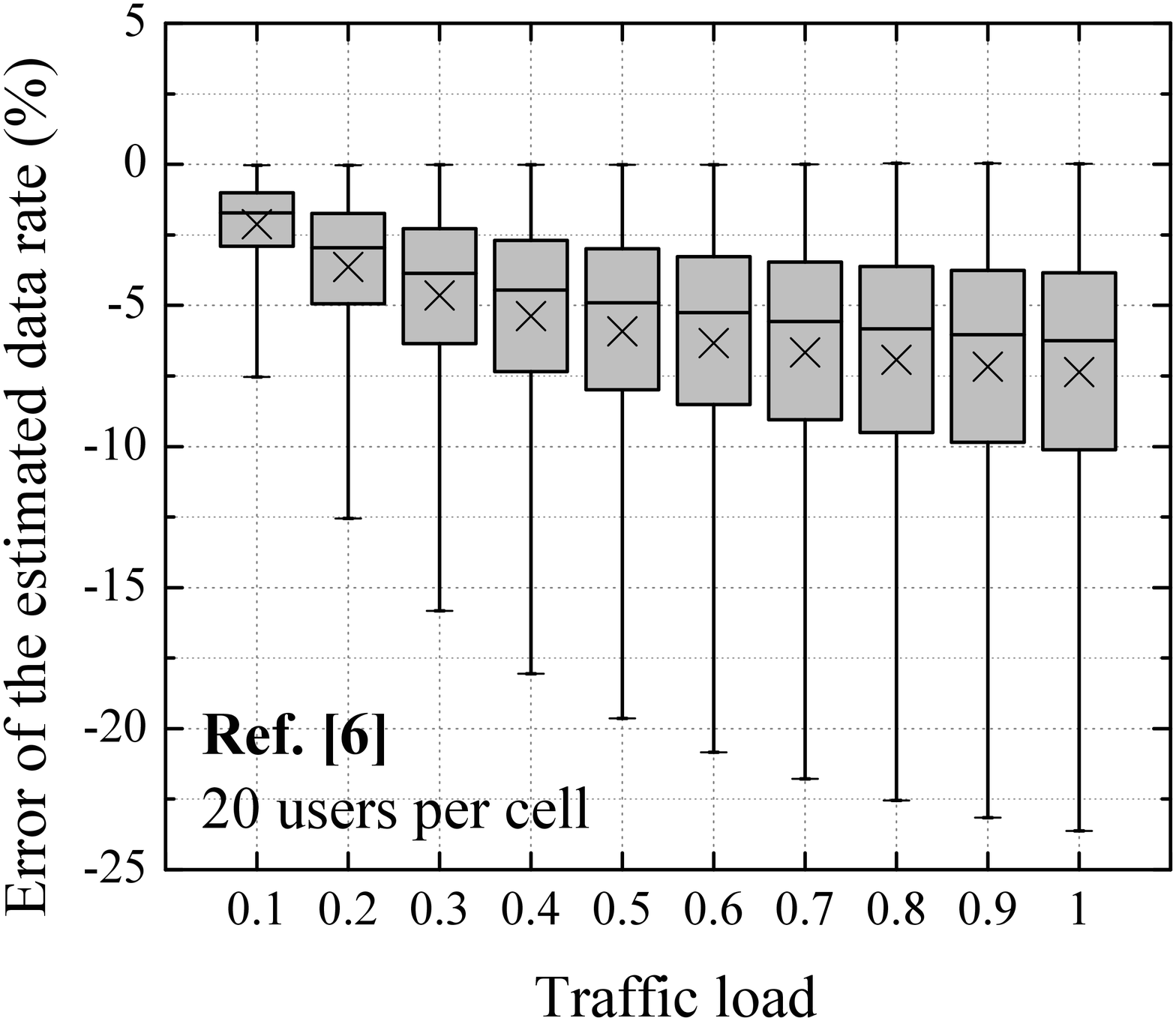}
 \caption{Error of estimated data rate with \protect\\  the model in [6].}
\label{IaN20UserBurst}}
\end{minipage}
\end{figure*}
%%%%%%%%%%%%%%%%%%%%%%%%%%%%%%%%%%%%%%%%%%%%%%%%%%%%%%%%%%%%%%%%%%%%%%%%%%%%%%%%%%%%%%%%%%%%%
Denoting the integral in \eqref{eqGAfullgain} as ${L\left( {N} \right)}$, i.e.,
\begin{equation}
L\left( N \right) = \int_0^1 z d{\left[ {{F_{\left( {0,1} \right)}}\left( z \right)} \right]^N},
\end{equation}
\noindent the estimated performance gain in \eqref{eqGAfullgain} is rewritten as
\begin{equation}\label{GAfull}
\tilde G_u \left( {\bf{U}}_b \right) = 1 + \frac{{{\sigma _{{u}}}}}{{{{ r}_u}}}L\left(  {\left| {\bf{U}}_b \right|}\right).
\end{equation}
\par
Substituting this into \eqref{Rburst}, we solve the closed-form expression of user throughput as
\begin{align}\label{GAburst}
 &{{\tilde R}_u}\left( {{\bf{U}}_b,\rho } \right) = {r_u}\rho \sum\limits_{u \in \left({\bf{V}} \subseteq {\bf{U}}_b\right)} {\left[ {\frac{{{{\tilde G}_u}\left( {\bf{V}} \right)}}{{\left| {\bf{V}} \right|}}{\rho ^{\left| {\bf{V}} \right| - 1}}{{\left( {1 - \rho } \right)}^{\left| {\bf{U}}_b \right| - \left| {\bf{V}} \right|}}} \right]}  \notag\\
 & = {r_u} \sum\limits_{n = 1}^{\left| {\bf{U}}_b \right|} {\left[ {\left( {\begin{array}{*{20}{c}}
   {\left| {\bf{U}}_b \right| - 1}  \\
   {n - 1}  \\
\end{array}} \right)\frac{{1 + {\sigma _u}r_u^{ - 1}L\left( n \right)}}{n}{\rho ^{n}}{{\left( {1 - \rho } \right)}^{\left| {\bf{U}}_b \right| - n}}} \right]}  \notag\\
 & = \frac{{{r_u}}}{{\left| {\bf{U}}_b \right|}}\left[ {1 - {{\left( {1 - \rho } \right)}^{\left| {\bf{U}} _b\right|}}} \right] + \frac{{{\sigma _u}}}{{\left| {\bf{U}}_b \right|}}l\left( {\left| {\bf{U}} _b\right|,\rho } \right),
\end{align}
\noindent where $l\left( {N,\rho } \right)$ represents
\begin{equation}
l\left( {N,\rho } \right) = \sum\limits_{n = 1}^N {\left[ {\left( {\begin{array}{*{20}{c}}
  N  \\
   n  \\
\end{array}} \right){\rho ^n}{{\left( {1 - \rho } \right)}^{N - n}}L\left( n \right)} \right]} .
\end{equation}
\noindent Specially, under saturated traffic, i.e., $\rho = 1$, we have
\begin{equation}
l\left( {N,1} \right) = L\left( N \right).
\end{equation}
\par
The average data rate with RR scheduling is calculated as
\begin{equation}\label{rateRR}
R_u^* \left( {{\bf{U}}_b,\rho } \right)= \frac{{{r_u}}}{{\left| {\bf{U}}_b \right|}}\left[ {1 - {{\left( {1 - \rho } \right)}^{\left| {\bf{U}}_b \right|}}} \right].
\end{equation}
\noindent According to \eqref{GAburst} and  \eqref{rateRR}, we obtain the performance gain of PFS over RR under bursty on-off traffic as
\begin{equation}\label{gainGA}
\tilde g_u \left( {{\bf{U}}_b,\rho } \right) = 1 + \frac{{{\sigma _u}}}{{{r_u}}}l\left( {\left| {\bf{U}}_b \right|,\rho } \right){\left[ {1 - {{\left( {1 - \rho } \right)}^{\left| {\bf{U}}_b \right|}}} \right]^{ - 1}}.
\end{equation}
\noindent Specially, while $\rho =1 $, i.e., under saturated traffic,
\begin{equation}
\tilde g_u\left( {{\bf{U}}_b,1} \right) = \tilde G_u\left( {\bf{U}} _b\right).
\end{equation}
\par
%\subsection{Evaluation of the GA-based Estimation}
We use system-level simulations to evaluate the accuracy of the GA-based data rate estimation. The network configurations are identical with those in~\cite{SECON2015}, and the scenario is analogous to the urban area of Berlin. The average number of user terminals per cell is 20 and they are uniformly randomly distributed over the cell areas. We set $\alpha = 1.5$ and $\beta = 10 \,s$ for the \emph{on} state, and thus $D_{on} = 30\,s$. $\lambda$ is set according to the test traffic load $\rho$. The error of the GA-based data rate estimation is presented in Fig.~\ref{GA20UserBurst}. When the traffic load is low, the estimation error is very small. However, it increases significantly with the traffic load. We calculate the estimation error with various numbers of users per cell under saturated traffic as shown in Fig.~\ref{GANUserFull}. The GA-based approach results in larger deviation when there are more users. The chance of \emph{on} state is large when the traffic load is high in the busty traffic scenario. Therefore, more users are likely to be active simultaneously and the estimation error increases in Fig.~\ref{GA20UserBurst}. The error of the estimated data rate by using the model in \cite{ACMTran2005} is also presented in Fig.~\ref{IaN20UserBurst}. This model results in worse estimation performance that has both an underestimation bias and much larger deviation compared to the GA-based analysis. %Therefore, it is not applicable for multi-cell scenario as GA.
\par
\section{Analysis of PFS Using Hybrid Approximation}
In order to remedy the shortcomings of GA under heavy traffic load and improve the estimation accuracy in multi-cell networks, we use the more accurate analytical model in~\cite{SECON2015}. This model is developed based on the multi-interference analysis and outperforms GA in terms of estimation accuracy under saturated traffic. Different from the GA model, MIA calculates the performance gain of a subset ${\bf{V}}$ considering the specific users in it instead of only the number of users. Therefore, it cannot be extended directly for bursty traffic analysis due to the combinatorial complexity problem that we explained after \eqref{Rburst}. We design a hybrid approximation (HA) by using MIA in the saturated traffic case ($\rho = 1$) and combining it with the GA-based estimation under bursty traffic ($\rho < 1$).
\par
%\subsection{MIA-based Data Rate Estimation}
Before formulating the HA model, we briefly introduce MIA. It considers multiple independent interference signals separately. The CDF and probability distribution function (PDF) of instantaneous user SINR under Rayleigh fading channel are derived in \cite{SECON2015} as
\begin{align}\label{eqCDFSINR}
{F_{{\Phi _u}}}\left( \phi  \right) =& {\rm{P}}\left\{ {{\Phi _u} < \phi } \right\} \notag\\
=& 1 - \exp \left[ { - {{\phi {\sigma _N}} \over {{p_{u,b}}}}} \right]\prod\limits_{i \in {{\bf{I}}_u}} {{{\left( {{{{p_{u,i}}} \over {{p_{u,b}}}}\phi  + 1} \right)}^{ - 1}}} ,
\end{align}
\begin{align}\label{eqPDFSINR}
{f_{{\Phi _u}}}\left( \phi  \right) =& \left[ {1 - {F_{{\Phi _u}}}\left( \phi  \right)} \right]\left[ {{{{\sigma _N}} \over {{p_{u,b}}}} + {\sum _{i \in {{\bf{I}}_u}}}{{\left( {\phi  + {{{p_{u,b}}} \over {{p_{u,i}}}}} \right)}^{ - 1}}} \right],\notag\\
& \phi>0.
\end{align}
\noindent Based on this stochastic SINR model, we calculate the average user data rate under saturated traffic with MIA as
\begin{align}\label{eqMIAR}
&{{\overline R}_u}\left( {\bf{U}}_b \right) = \notag \\&\int_0^\infty  {r\left( \phi  \right)} {f_{{\Phi _u}}}\left( \phi  \right)\prod\limits_{v \in \left({\bf{U}}_b/u\right)} {{F_{{\Phi _v}}}\left( {{r^{ - 1}}\left( {\frac{{r\left( \phi  \right){{\overline R}_v}\left( {\bf{U}}_b\right)}}{{{{\overline R}_u}\left( {\bf{U}} _b\right)}}} \right)} \right)d\phi },
\end{align}
\noindent where ${r\left( \phi  \right)}$ is the data rate mapping function which is based on the Shannon capacity as in the GA-based estimation~\cite{WCNC2008}.
%\subsection{Hybrid Approximation}
\par
%%%%%%%%%%%%%%%%%%%%%%%%%%%%%%%%%%%%%%%%%%%%%%%%%%%%%%%%%%%%%%%%%%%%%%%%%%%%%%%%%%%%%%%%%%%%%
\newcounter{TempEqCnt}
\setcounter{TempEqCnt}{\value{equation}}
\setcounter{equation}{24}
\begin{figure*}[t]
% ensure that we have normalsize text
\normalsize
\begin{equation}\label{HA}%\tag{c}
{{\hat g}_u}\left( {{\bf{U}}_b,\rho } \right) = 1 + \left( {1 - \rho } \right){\tilde\eta _u}\left( {{\bf{U}}_b,\rho } \right) + \rho \underbrace {{{{\tilde \eta _u}\left( {{\bf{U}}_b,\rho } \right)} \over {{\tilde\eta _u}\left( {{\bf{U}}_b,1} \right)}}}_{\left( a \right)}\underbrace {\left[ {{{\overline G}_u}\left( {\bf{U}}_b \right) - 1} \right]}_{\left( b \right)}.
\end{equation}
\hrulefill
% The spacer can be tweaked to stop underfull vboxes.
\vspace*{-5pt}
\end{figure*}
\setcounter{equation}{\value{TempEqCnt}}
%%%%%%%%%%%%%%%%%%%%%%%%%%%%%%%%%%%%%%%%%%%%%%%%%%%%%%%%%%%%%%%%%%%%%%%%%%%%%%%%%%%%%%%%%%%%%
By combining the GA- and MIA-based approaches, we design a hybrid approximation (HA) as follows. We denote the increment part of the estimated performance gain in \eqref{gainGA} as ${\tilde \eta _u}\left( {{\bf{U}}_b,\rho } \right)$, i.e.,
\begin{equation}\label{gainGA2}
\tilde g_u \left( {{\bf{U}}_b,\rho } \right) = 1 + {\tilde \eta _u}\left( {{\bf{U}}_b,\rho } \right).
\end{equation}
\par
We consider a hybrid strategy which uses the GA-based results under low traffic load and the MIA-based ones under high traffic load. In this way, the inaccuracy of GA can be remitted while the number of active users is high. Thus, the proposed HA model is formulated as in \eqref{HA}, where ${{\overline G}_u}\left( {\bf{U}}_b \right)$ is the performance gain calculated by MIA under saturated traffic, i.e.,
\setcounter{equation}{25}
\begin{equation}\label{eqMIAGain}
{{\overline G}_u}\left( {\bf{U}}_b \right) = {{\overline R}_u}\left( {\bf{U}} _b\right)/R_u^*\left( {\bf{U}}_b ,1\right).
\end{equation}
In \eqref{HA}, the performance increment estimated by GA, i.e., ${\tilde \eta _u}\left( {{\bf{U}}_b,\rho } \right)$, is weighted by $\left( {1 - \rho } \right)$. Item (a) is a ratio of the performance increment between unsaturated and saturated traffic, while item (b) is the performance increment estimated by MIA under saturated traffic. The combination of (a) and (b) is weighted by $\rho$.
\par
Specifically, we have the limiting values as
\begin{align}\label{eqboundHA}
 {\left. {{{\hat g}_u}\left( {{\bf{U}}_b,\rho } \right)} \right|_{\rho  \to 0}} &= {{\tilde g}_u}\left( {{\bf{U}}_b,\rho } \right), \quad{\rm{and}}  \\
 {{\hat g}_u}\left( {{\bf{U}}_b,1} \right)& = {{\overline G}_u}\left( {\bf{U}}_b \right).
\end{align}
\noindent Thus, HA yields identical results under extremely low and high traffic load with GA and MIA models, correspondingly.
\par
%%%%%%%%%%%%%%%%%%%%%%%%%%%%%%%%%%%%%%%%%%%%%%%%%%%%%%%%%%%%%%%%%%%%%%%%%%%%%%%%%%%%%%%%%%%%
\begin{figure}
\centering
{\includegraphics[width=3.5in]{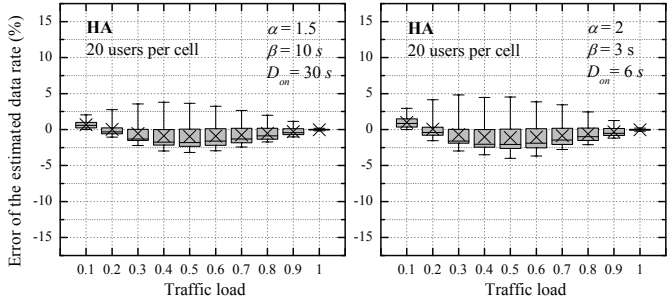}}
\caption{Error of the estimated data rate with HA under bursty on-off traffic.}% (20 users per cell).}
\label{HA20UserBurst}
\end{figure}
%%%%%%%%%%%%%%%%%%%%%%%%%%%%%%%%%%%%%%%%%%%%%%%%%%%%%%%%%%%%%%%%%%%%%%%%%%%%%%%%%%%%%%%%%%%%%
%%%%%%%%%%%%%%%%%%%%%%%%%%%%%%%%%%%%%%%%%%%%%%%%%%%%%%%%%%%%%%%%%%%%%%%%%%%%%%%%%%%%%%%%%%%%%
\begin{figure}
\centering
{\includegraphics[width=3.5in]{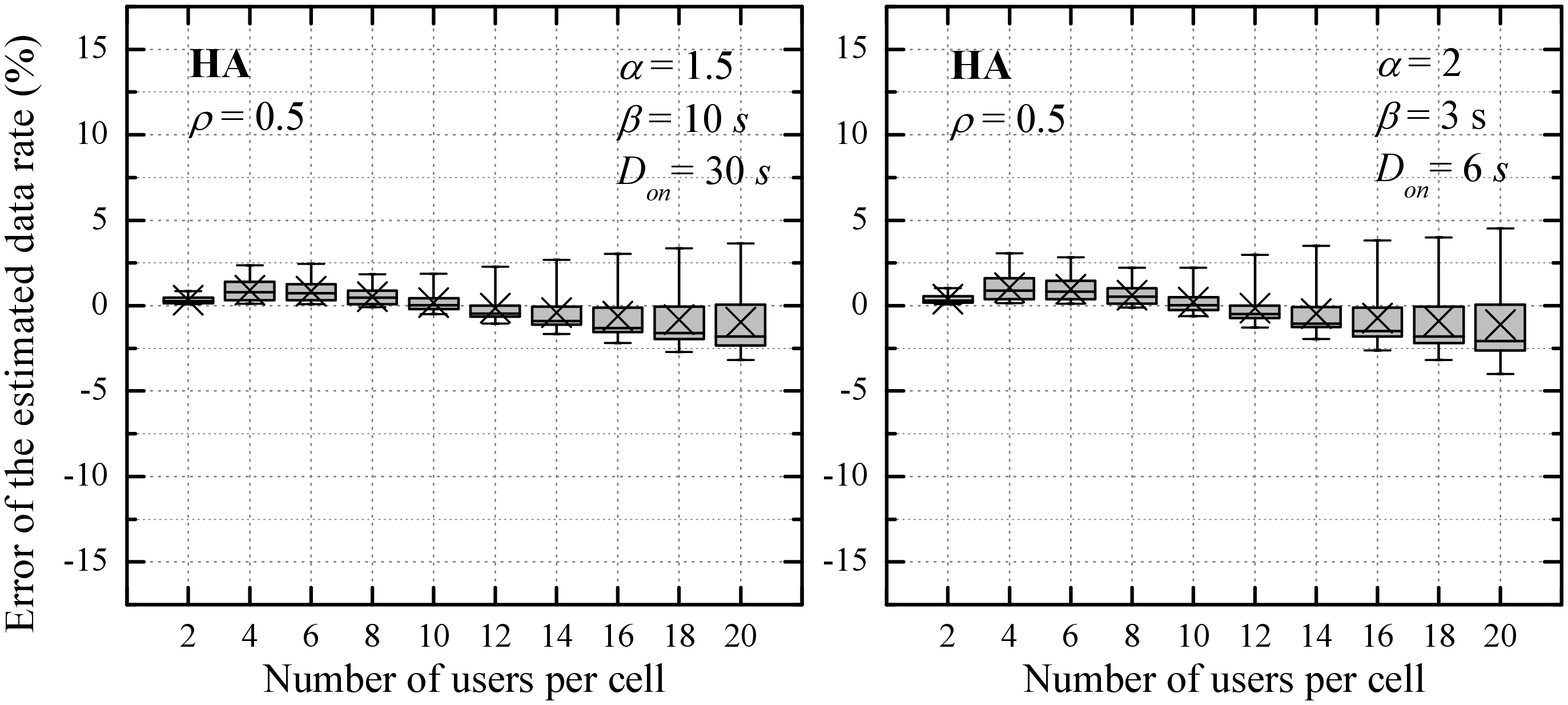}}
\caption{Error of the estimated data rate with various numbers of users.}% ($\rho = 0.5$).}
\label{HANUserBurst50}
\end{figure}
%%%%%%%%%%%%%%%%%%%%%%%%%%%%%%%%%%%%%%%%%%%%%%%%%%%%%%%%%%%%%%%%%%%%%%%%%%%%%%%%%%%%%%%%%%%%%
The simulation results of our HA model are shown in Fig.~\ref{HA20UserBurst}. It achieves significant improvement in terms of data rate estimation accuracy, especially under high traffic load. In addition, the HA model is tested with more transient session periods, i.e., $D_{on}=6\,s$. In comparison with the case where $D_{on} = 30\,s$, the estimation errors vary only slightly, indicating that the accuracy of the HA model is not sensitive to the change of session duration. We further investigate the estimation accuracy with $\rho=0.5$ since the estimation error is larger under median traffic load as shown in Fig.~\ref{HA20UserBurst}. The simulation results with various numbers of users are presented in Fig.~\ref{HANUserBurst50}. %When there are fewer users in the network, most of the estimated user rates are slightly larger than the actual ones due to the overestimation caused by GA in this case which can also be observed in Fig.~\ref{GANUserFull}.
The estimation error is always within $\pm 5\%$ with various numbers of users per cell, which is a significant reduction in comparison with the pure GA-based approach. Thus, the HA model is much more accurate for practical applications.
\par
%Hybrid approximation:
%\begin{align}\label{HA}
% {{\hat g}_u}\left( {{\bf{U}},\rho } \right) =& 1 + \left\{ {\rho \left[ {\frac{{{{\overline g}_u}\left( {{\bf{U}},1} \right) - 1}}{{{{\tilde g}_u}\left( {{\bf{U}},1} \right) - 1}}} \right]\left[ {{{\tilde g}_u}\left( {{\bf{U}},\rho } \right) - 1} \right] + \left( {1 - \rho } \right)\left[ {{{\tilde g}_u}\left( {{\bf{U}},\rho } \right) - 1} \right]} \right\} \\
%  =& 1 + l\left( {\left| {\bf{U}} \right|,\rho } \right)\left\{ {\rho \frac{{\left[ {{{\overline G}_u}\left( {\bf{U}} \right) - 1} \right]}}{{L\left( {\left| {\bf{U}} \right|} \right)}} + \left( {1 - \rho } \right)\frac{{{\sigma _u}}}{{{r_u}}}} \right\} \notag
%\end{align}
\section{Conclusions}\label{SecConclu}
In this letter, we derived the first analytical solution for estimating user data rates of PFS under bursty on-off traffic. We used Gaussian approximation to solve the closed-form expression of user data rates. The simulation results show that GA-based analysis is accurate under low traffic load. However, the estimation accuracy declines significantly as the number of active users increases. In order to improve the accuracy of data rate estimation in multi-cell network, we developed a hybrid approximation by employing MIA combined with GA as the traffic load increases. The simulation results verify that this approach increases the estimation accuracy significantly. The errors of the analytical results are lower than $5\%$ and are shown to be insensitive to changes in session duration, traffic load and user density.
\par

% biography section
%
% If you have an EPS/PDF photo (graphicx package needed) extra braces are
% needed around the contents of the optional argument to biography to prevent
% the LaTeX parser from getting confused when it sees the complicated
% \includegraphics command within an optional argument. (You could create
% your own custom macro containing the \includegraphics command to make things
% simpler here.)
%\begin{IEEEbiography}[{\includegraphics[width=1in,height=1.25in,clip,keepaspectratio]{mshell}}]{Michael Shell}
% or if you just want to reserve a space for a photo:
%
%\begin{IEEEbiography}{Michael Shell}
%Biography text here.
%\end{IEEEbiography}
%
%% if you will not have a photo at all:
%\begin{IEEEbiographynophoto}{John Doe}
%Biography text here.
%\end{IEEEbiographynophoto}
%
%% insert where needed to balance the two columns on the last page with
%% biographies
%%\newpage
%
%\begin{IEEEbiographynophoto}{Jane Doe}
%Biography text here.
%\end{IEEEbiographynophoto}

% You can push biographies down or up by placing
% a \vfill before or after them. The appropriate
% use of \vfill depends on what kind of text is
% on the last page and whether or not the columns
% are being equalized.

%\vfill

% Can be used to pull up biographies so that the bottom of the last one
% is flush with the other column.
%\enlargethispage{-5in}

% that's all folks

\end{document}